\documentstyle[12pt]{article}
\newcommand{\ber}{\begin{eqnarray}}
\newcommand{\eer}{\end{eqnarray}}
\newcommand{\bea}{\begin{equation}}
\newcommand{\eea}{\end{equation}}
\begin{document}
\title{Codon Distributions in DNA} 
\author{$A.Som^{1,a},\: S.Chattopadhyay^{1,b},\: J.Chakrabarti^{1,c}\: and \: D.Bandyopadhyay^{2,d}$\\\\
$^1$ Department of Theoretical Physics \\
  Indian Association for the Cultivation of Science \\
  Calcutta 700 032,  INDIA \\\\
$^2$ Linguistics Research Unit \\
Indian Statistical Institute \\
Calcutta 700 035,  INDIA }
\date{}
\maketitle
\newpage
\begin{abstract}
The codons, sixtyfour in number, are distributed over the coding parts of DNA sequences. The distribution function is the plot of frequency-versus-rank of the codons. These distributions are characterised by parameters that are almost universal, i.e., gene independent. There is but a small part that depends on the gene. We present the theory to calculate the universal (gene-independent) part. The part that is gene-specific, however, has undetermined overlaps and fluctuations.
\\\\
\par\noindent
PACS number(s): 87.10.+e, 85.15.-v, 05.40.+j, 72.70+m
\\\\
\par\noindent
E-mails: \\
$^{a}$   tpas@mahendra.iacs.res.in \\
$^{b}$   tpsc@mahendra.iacs.res.in \\
$^{c}$   tpjc@mahendra.iacs.res.in \\
$^{d}$   bandyo@isical.ac.in \\
\\\\
\end{abstract}

\newpage
\section{Introduction}
The methods of statistical linguistics are used in recent  years to study DNA sequences[1]. The genome projects generate large volumes of data  on DNA. Fast and reliable computational tools to analyse this huge data of billlions of bases  are required. The idea is to identify features in the sequences and to correlate them with known biological functions. The methods of statistical linguistics[2] could provide reliable computational algorithms. This is what we investigate here.
\par
The sequences are made of the nucleotide bases A, C, G and T. The arrangement of the bases over the linear chain determines all the information there is in DNA. The regions that code for proteins, the coding regions (or the exons), have bases working in groups of three to make proteins. These triplets are called codons. The biologically meaningful words are these codons. The noncoding parts consist of the introns and the flanks. These are presumed important in regulatory and promotional activities. The biologically meaningful word structures in these regions are not known. A gene generally comprises of a number of exon regions separated by introns. Since the biological functions thus far are associated with the triplet codons, we concern ourselves only with these triplet words, the codons. Therefore, in our analysis, instead of an entire gene, we consider the coding DNA sequence (CDS) region of the gene, where the exon segments are put together, splicing the introns out. 
\par
Natural languages are characterised by structures determined by rules of grammar. The words put together with these rules carry sense. The rules give coherence and meaning to long texts. The languages have this long-range order. The frequency spectra show the presence of the long periods. These are identified by the $\frac{1}{f^{\beta}}$ type behaviour in the low frequency region[3]. Words placed at random will have quite different frequency spectrum with no long-range behavior. The early work on natural languages dealing with the statistical distributions of words, done by Zipf [4], assigned ranks to the words. The word most frequent has rank=1; the next most has rank=2 and so on. Zipf showed that for natural languages the plot of frequency, $f_n$, versus rank, n, is of the power-law form:
\bea
f_n = \frac{f_{1}}{n^{\alpha}} 
\eea
 where $f_{1}$ is the frequency of rank 1. In the Zipf's original analysis the power-index $\alpha$ was assumed to be one. Subsequent studies have allowed for deviations from one.
\par
 The DNA sequence of the letters A, C, G and T does have $\frac{1}{f^{\beta}}$ frequency spectrum[5]. It is possible, therefore, that the sequences have long-range order and underlying grammer rules. The opinion on this issue remains divided[6]. Some have taken the view that DNA is language-like[7]. In the coding regions the long periods have lower incidence than in the non-coding parts. The Zipf-type fits in DNA regions (with overlapping n-tuples) have shown that the index $\alpha$ is higher in the non-coding segments over the coding ones. The averaged $\alpha$ over several overlapping n-tuples is nearer to the value for natural languages for non-coding segments than the coding ones[1,7].
\par
 The body of evidence presented in support of the language-like features of DNA has remained ambiguous[8]. For one it is not known how the power-law Zipf-behaviour of natural languages is connected to the long-range correlations[9]. It is known, for instance, that pseudorandom sequences satisfy Zipf- behaviour. Further, it is known that the frequencies of A, C, G and T vary somewhat more for the introns and the flanks over the exons[10]. The $\lq\lq$long-range" order that is observed for these noncoding regions may be an  outcome of the frequency differences. The higher value of the Zipf index for the noncoding segments may again be  ascribed to these differences in the frequencies of the bases.
\par
The importance of statistical linguistics as a computational tool remains insufficiently explored for DNA sequences. While the Zipf law is probably not connected to the deeper features of languages such as the universal grammar, the coherence and the long periods, it could still be useful. For instance,  the index $\alpha$ of languages could be (and is) used in computer algorithms to identify authors. The texts generated by authors vary slightly in their Zipf index. The index, therefore, identifies the author. Could one use similar algorithms to identify regions from the genome segments and relate them to their biological functions?
\par
As precision and reliability are important we have weighed the merits of power-law fits over exponential fits. Since we are solely concerned with non-overlapping 3-tuples (i.e. the codons), we find the exponential fits have consistently lower $\chi^{2}$. [Chi-square ($\chi^{2}$) is the sum of the ratio of the squared difference between observed value at the i$^{th}$ point (o$_i$) and the expected value at the i$^{th}$ point (e$_i$) to the expected value at the i$^{th}$ point (e$_i$), i.e., $\chi^{2}$ = $\sum_{i}\frac{(o_{i}-e_{i})^2}{e_{i}}$; where the sum i runs over the number of points of the fit. The value of $\chi^{2}$ depends on the total number of points to be fit minus one, sometimes called the degree of freedom, df.] The exponentials, therefore, provide better fits. That the power-law fits for DNA sequences are worse than the exponentials have also been observed by others[11]. The power law of Zipf is characterised by two parameters, the index $\alpha$ and the frequency of rank one, i.e. $f_{1}$. The number of parameters for the exponential fit is of interest to us. The Zipf's law is used to find the relationship connecting vocabulary to the text-length. Such connection does exist for the exponential fit as well.
\par
 The parameters of the exponential rank-frequency relation depend crucially on the text-length. Once this parameter is known, the approximate length of the segment gets known as well. Indeed, the exponential fits are largely determined by two quantities, the frequency of rank 1, i.e., $f_{1}$ and  the text-length of the sequence. There is however a small part that is characteristic of the gene. This signature of the gene is potentially useful in generating algorithms to identify the gene and relate to the biological functions.

\section{The Approach}
Out of the four bases A, C, G $\&$ T we have 4 $\times$ 4 $\times$ 4 = 64 possible triplets. Three combinations, namely, TAA, TAG $\&$ TGA are the stop condons. Thus 64 - 3=61 is the meaningful vocabulary. The codon most frequent has rank n=1, the next most has n=2 and so on. We define frequency, $f$, of a particular codon as the number of times it appears in the sequence. [Note this definition is different from some of the references where $f_{n}$=$\frac{Number\:of\:words\:of\:rank\:n}{Total\:number\:of\:words}$. The frequency of rank n is $f_{n}$. Here both frequency($f$) and rank(n) are dimensionless.
\par
 Observations on the CDS reveal that many codons may have the same frequency. Note that the CDS we are dealing with are relatively short sequences of several hundred to several thousand bases. This problem of multiple codons having the same frequency is called frequency degeneracy.
\par
First, as we consider only codons, 61 in number, the problem of saturation of vocabulary for large text-length is clear. However, for most genes we observe that the actual usage of codons is smaller than 61. The codon usage is sometimes referred to as the vocabulary, i.e. the total number of different codons, used in the CDS.
\par
From the Zipf's law [equation(1)] with $\alpha$=1 we have 
$$ ln (f_{n}) = ln (f_{1}) - ln (n) $$
If we plot ln($f_{n}$) vs ln(n) we have a  straight line with slope -1 and intercept on the y-axis at ln($f_{1}$). Clearly, the maximum rank is just equal to $f_{1}$. When $\alpha$ deviates from 1, $f_{1}$ and the maximum rank are connected to each other through  $\alpha$. The maximum rank (i.e. the vocabulary) along with $f_{1}$ (or $\alpha$) determine the text-length l, i.e., the total number of triplets, as follows :
$$ l= f_1+f_2+f_1+f_3+ .... +f_n $$
$$  = f_1(1+ \frac{1}{2^\alpha}+ \frac{1}{3^\alpha}+  .... +\frac{1}{n^\alpha})$$
Thus, $\alpha$ may be thought of as a function of  $f_{1}$  and the text-length l. We want to arrive at the corresponding relation for our exponential fits.

\section{The Exponential Fit}
All the degenerate frequencies are assigned  different rank number. Thus if CCG and CAG have the same frequency of occurrence they belong to two different ranks (one following the other) in our work. Therefore, here too, the codon usage, maximum rank and vocabulary are synonymous. The exponential function that connects frequency to rank is 
\bea
f_{n} = f_{1}exp\{-\beta(n-1)\}
\eea
where $\beta$, a dimensionless constant for a particular gene, is to be determined from the fit.
\par
We have tried this fit function on over 300 CDS. The CDS are sourced from the EMBL[12] and the GenBank[13] data bases. Table 1 gives the values of $\beta$ for some of the sequences under study. The plots showing the fit is figure(1). 
\par
The index $\beta$ in the exponential of equation(2) takes different values for the genes. It turns out, however, that $\beta$ is not completely a free parameter. Indeed, from Table 1, we notice that CDS that have text-lengths and also $f_1$ that are close have similar, though not identical, $\beta$ values. Notice, for instance, the $\beta$-globin CDS from the chicken and the clawed frog have the same l and $f_1$, 147 and 9 respectively; whereas the lysozyme CDS from the fish, $\it{Cyprinus\:carpio}$ has 146 as l and 9 as $f_1$. The $\beta$ values for the $\beta$-globin CDS of the chicken and the frog are 0.05773 and 0.05772; while the lysozyme CDS, though functionally quite unrelated to the $\beta$-globin, has the $\beta$ value of 0.06056. So the value of $\beta$ is determined to a considerable extent by $f_{1}$ and the text-length of the sequence, l. There is but a part in $\beta$ that is characteristic of the gene.

\section{Plot of $\beta$ vs. $f_{1}$}
Figure(2) gives plots of $\beta$ vs $f_{1}$ for four complete CDS coding for $\alpha$-globin, $\beta$-globin, phosphoglycerate kinase and globulin proteins. The $\chi^{2}$ values indicate that the relationship between $\beta$ and $f_{1}$ is linear to a good approximation. The plot for each CDS involves data on the gene from different species. These are sourced from GenBank.
Each of the linear plots are specific to the gene. The evolution of the genes, as we move higher in the evolutionary hierarchy, does not significantly alter the overall text-length of the CDS regions. 

\par
The slope of the globin CDS, the $\alpha$ and the $\beta$, are nearly equal. As we show in the subsequent pages the value of $\beta$ is considerably determined by $f_{1}$ and l. There is but a small part that is unique to the gene. For the case of the $\alpha$ and the $\beta$ globins notice that the text-lengths of these CDS vary in a small range between 143 and 147. Table 1 shows that any two quite unrelated CDS can have $\beta$ values that are close provided their text-lengths and the $f_{1}$ are nearly equal. 

\par
The plots in figure(3) of $\beta$ vs $f_{1}$ keep the text-length l fixed at 140 for the same four genes. Though the closeness in the values of the slope indeed show the influence of l on the $\beta$ value, the small differences indicate the presence of the l-independent part in the $\beta$ value. 

\par
That the $\beta$ values are not completely determined by 
$f_{1}$ and l, but do have a component, albeit small, coming from the genes is illustrated in our next plot, figure(4).
A number of different CDS, each from a different organism, were chosen and cut at three different text-lengths 30, 140 and 300, i.e., we considered only the first 30, 140 and 300 triplets respectively out of the whole CDS. The plot of $\beta$ vs $f_{1}$ for these three different 
text-lengths indicates that when the text-length is held fixed, but the genes are varied, the exponential gives a better fit over the linear. It is noteworthy that even though the genes are unrelated in as far as their biological functions are concerned, the codon distributions, described by the experimental fit of figure(4), are not completely unrelated.
\par
Taken together, the two plots, figure(3) and figure(4), tell us: \\
(i) When the text-length, l, is held fixed, and the genes are not varied, the plot of $\beta$ vs $f_{1}$ is linear and \\
(ii) When the text-length, l, is held fixed, and the genes are varied, the plot of $\beta$ vs $f_{1}$ is exponential. 
\par
Thus, we conclude that the value of $\beta$ does have a part that is gene specific.

\section{Plot of $\beta$ vs l}
$\beta$, as we have observed from Table 1, depends on $f_{1}$ and l. Beyond that there is the part that is gene specific. In other words the parameters of the functional fit do depend, in a small way, on the gene. This dependence we discuss later. Here, in this section, we concern ourselves with the dependence of $\beta$ on the text-length of the CDS. 
\par
We plot $\beta$ vs l keeping $f_{1}$ fixed. The plots in figure(5) show the dependence for four different values of $f_{1}$, namely $f_{1}$=7, $f_{1}$=9, $f_{1}$=20, and $f_{1}$=38.
\par
In plotting figure(5) we considered the $f_{1}$ values of the natural CDS. We had the option to cut the CDS into fragments to suit our value of $f_{1}$. This procedure turned out to be arbitrary as the $f_{1}$ value may remain fixed over some hundred bases. Cutting into fragments is nonunique. It was, therefore, difficult to restrict our study of $\beta$ vs l for a particular gene. For a specific CDS (from different species) the text-length does not vary significantly in most cases. Therefore for a fixed value of $f_{1}$ the CDS were searched over different genes. Thus $f_{1}$ is held fixed, but genes vary.
\par
Though more data for each gene could have improved the result, nevertheless the relationship between $\beta$ and l for fixed $f_{1}$ has a linear trend. As the text-length increases $\beta$ decreases. However, the plots for different values of $f_{1}$ are not parallel. They depend on $f_{1}$. The slope reaches a maximum at around $f_{1}$ = 10 and tend to decrease as we go away from $f_{1}$=10 on either side. For large values of $f_{1}$, the slopes tend to become parallel. 

\section{Theory of $\beta$}
We have seen $\beta$ depends on the text-length, l, and the frequency of rank 1, $f_{1}$. \\
(1) When the text-length l is held fixed, genes not varied, $\beta$
depends linearly on $f_{1}$. The plot of $\beta$ vs $f_{1}$ shows that $\frac{\Delta{\beta}}{\Delta{f_{1}}}$ is positive.\\
(2) When the text-length is kept fixed, but the genes are varied, the plot of $\beta$ vs $f_{1}$ show deviations from linearity. An exponential fit appears more appropriate. \\
(3) When $f_{1}$ is held fixed (genes are varied as well) the plot of $\beta$
vs l shows an approximate linear behaviour. $\frac{\Delta{\beta}}{\Delta{l}}$ is negative. Note that, because of the points mentioned earlier, the variations in l (in figure 5) are over a rather small range. As a result the full l-dependence is not clear from figure(5).
\par
In this section we investigate $\beta$ theoretically. Let us denote the maximum rank by $n_{max}$. Since the frequency of $n_{max}$ is almost always one, we get 
\bea
1= f_{1} exp{\lbrace}-{\beta}(n_{max}-1){\rbrace}
\eea
Or, 
\bea
n_{max} = \frac{ln{f_1}}{\beta}+1
\eea
The text-length l is just the sum over all the frequencies. Thus,
\bea
l = \sum_{n=1}^{n_{max}}f_{1}.e^{-\beta(n-1)}
\eea
\bea
=\frac{f_{1}(1-e^{-\beta(n_{max}-1)})}{1-e^{-\beta}}
\eea
Substituting for $n_{max}$ from equation(4), we get 
\bea
l = \frac{f_{1}-1}{1-e^{-\beta}} 
\eea
Thus,
\bea
\beta = -ln[1-\frac{1}{l}(f_{1}-1)] 
\eea
Since, the quantity $\frac{f_{1}}{l}$ is small compared to one, we get, to
the first approximation 
\bea
\beta=\frac{f_1-1}{l} + higher\:\: orders  
\eea
Equation(9) tells us \\
(i) $\beta$ vs $f_{1}$, when l is kept fixed, is linear; the slope is positive.\\
(ii) $\beta$ vs l, with $f_{1}$ fixed, is hyperbolic. If the text-length variation is small we expect an approximate linear relation with negative slope (as observed in figure(5)). How good the relation(9) is checked in Table 1.
\par
While the relation(9) tells us that $\beta$ is entirely determined by the ratio of $f{_1}$-1 to l, figure(3) tells us that this quantity does have a characteristic dependence on the gene family. We conclude, therefore, that the relation(9) does not determine $\beta$ entirely. There is a part that is gene specific. The theoretical values of $\beta$, equation(9), is reasonably close to the values obtained from the CDS. The dependence of $\beta$ on $f{_1}$ and l of equation(9) is gene-independent. It is the universal part of  $\beta$. The deviation from this universal part, even though small, is established in figure(3) and figure(4). We define the quantity $\beta^{'}$ that gives a measure of this deviation through the relation:
\ber
\beta & = & [ \frac{f_{1}-1}{l} + \frac{1}{2}\frac{(f_{1}-1)^{2}}{l^{2}}]\beta^{'} \nonumber \\
 & = & \beta_{Th}\:\:.\:\:\beta^{'} 
\eer
where ${\beta_{Th} = [ \frac{f_{1}-1}{l} + \frac{1}{2}\frac{(f_{1}-1)^{2}}{l^{2}}]}$. \\
We have retained the first two orders in $\frac{f_1}{l}$ [ of equation(8)]. This is to make sure the higher-orders in $\frac{f_1}{l}$ do not account for the deviations. The values of $\beta{'}$ appear in the last columm of Table 1.

\section{$\beta$, $\beta^{'}$ and Evolution}
We get back to Table 1 for the CDS of $\alpha$-globin, $\beta$-globin, insulin and globulin. We notice the value of $f_{1}$ increases as we walk up along the ladder of evolution. The increase in $f_{1}$  increases $\beta$ while the text-length of the CDS does not change significantly in evolution. The results for insulin and the globulin CDS [Table 1] carry at least one exception. Interestingly, for both these CDS, the exceptional species is the same, the rabbit. The rabbit has $f_{1}$  and $\beta$ values greater than the human for these two CDS. The number of exceptions increase for the two globins. Some fishes show greater $f_{1}$ (and hence $\beta$) values than the amphibian species, the African clawed frog. If we average $\beta$ for the mammals we find it always exceeds the other groups. 
\par
On the other hand, if we compare the $\beta^{'}$ values for each of these four CDS, $\alpha$-globin and globulin do not show any clear pattern. In insulin, the $\beta^{'}$ values increase as we move from fish to mammals through amphibia. But the syrian hamster CDS is found to have lower $\beta^{'}$ than the clawed frog CDS. Besides the rat has greater $\beta^{'}$ compared to the human. In $\beta$-globin, the Atlantic salmon fish stands as an exception. Otherwise, the $\beta^{'}$ value increases from amphibia, bird to mammals. But here the representatives of amphibia and bird have the same value, and the lemur exceeds the value of human. We conclude that the value of $\beta^{'}$, though independent of l and f$_1$, is less species specific; whereas the value of $\beta$ does have evolutionary content.    

\section{Gene-Specific Signatures}
In figure(2) we showed that $\beta$ vs $f_{1}$ is a straight line when the genes are not varied. When the genes are varied, but the text-length is held constant, the relationship of $\beta$ to $f_{1}$ is no longer linear. The exponential fit is appropriate for this case. This led us to conclude that there is a part to $\beta$ that is gene-specific.
\par 
In figure(3) we plotted $\beta$ vs $f_1$ keeping the genes fixed for different organisms. The slope $\frac{\Delta{\beta}}{\Delta{f_{1}}}$ is  a characteristic of the gene. There is a variation in the slope as we go from one gene to another.
\par
The regular, namely exponential form, obtained in figure(4) in the plot of $\beta$ vs $f_1$, l being kept constant, tells us that the variations of $\beta$, as we go from one gene to another, is orderly.
\par
$\beta$ has a part that is gene independent. We isolate this universal component of $\beta$ theoretically. This part comes out to be a function of the text-length of the sequence and the frequency of rank 1, i.e. $f_1$. The quantity $\beta^{'}$, defined in equation(10), measures the deviation of the actual $\beta$ from this universal,  gene-independent, contribution given in equation(10). If the gene specific features are not dominant, $\beta^{'}$ should be close to one. Table 1 gives us the values of $\beta^{'}$. Clearly, the gene specific components in $\beta$ could be as high as 40$\%$ (as in insulin). We are led to conclude that the methods of statistical linguistics, of the Zipf variety, has the potential  in algorithms to identify genes from the databases.
\par
The quantity $\beta^{'}$ that isolates the gene-specific components of $\beta$ is however not unique to genes. Observations on $\beta^{'}$ (Table 1) show that the range of variations in $\beta^{'}$ do overlap for different genes. There continues to be undetermined fluctuations in the values of $\beta^{'}$. Work is currently in progress to isolate the unique gene-identifying signatures in the Zipf-approach.

\newpage\noindent
$ {\bf{Figure\:Legends}} $ \\\\
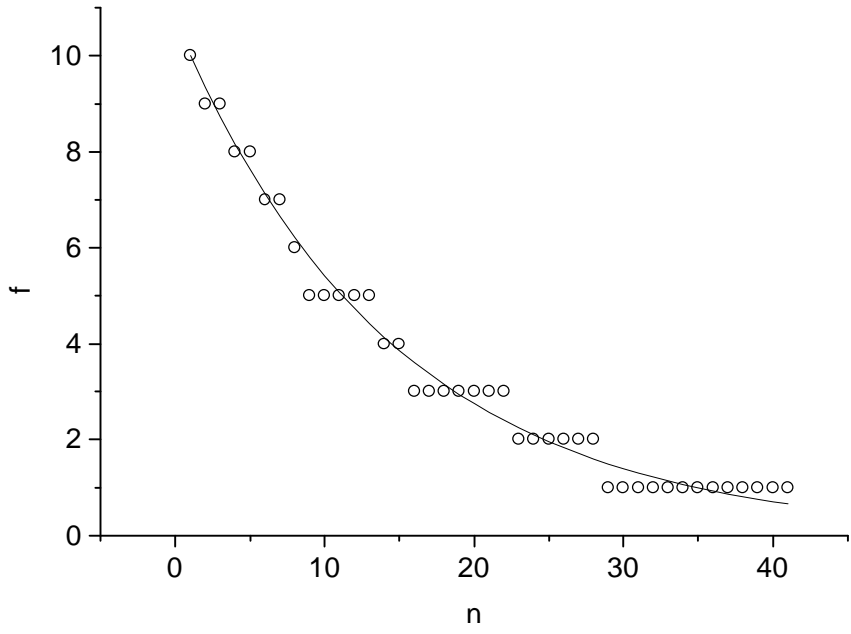
Figure 1. The plots of frequency (f) vs. rank (n) are the exponential functions (equation 2). Here different codons with the same frequency of occurrence are given consecutive ranks. The data corresponds to the $\alpha$-globin CDS from Duck (Acc. No. J00923). The $\beta$ value comes out to be 0.06801. The text-length, l of the CDS is 143; f$_1$ is 10.
\\\\
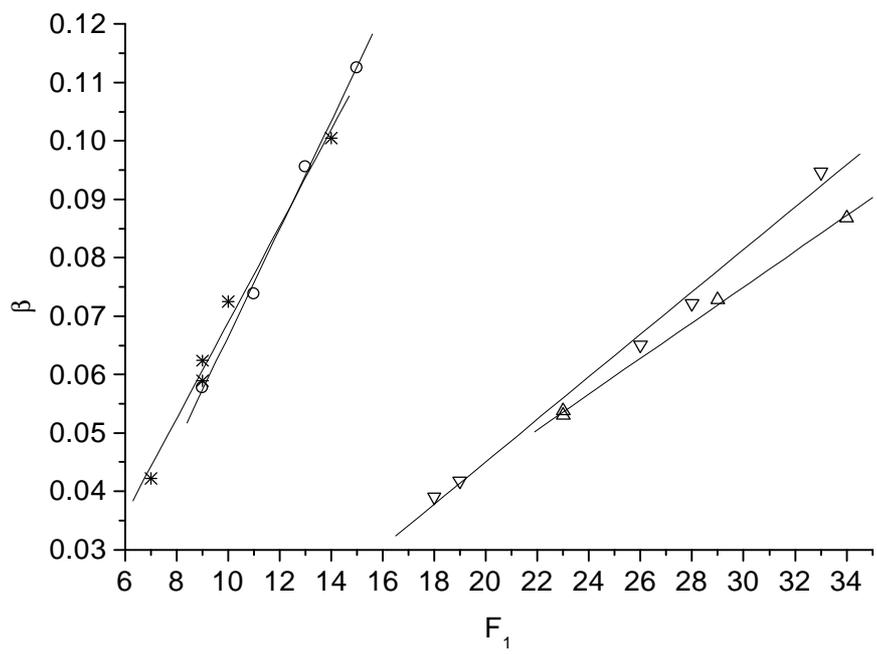
Figure 2. $\beta$ is plotted as a function of f$_1$ for the natural CDS of 4 different proteins from various species. The relationship turns out to be linear. 

\begin{table}[h]
\renewcommand{\arraystretch}{1.5}
\begin{tabular}{c c c c c c }
symbol & CDS & range of l & m & c & sd \\
\hline
$\star$ & $\alpha$-globin : & 142-151 & 0.0083 & -0.0136 & 0.0029 \\
$\circ$ & $\beta$-globin  : & 146-149 & 0.0092 & -0.0258 & 0.0014 \\
$\bigtriangleup$ & phosphoglycerate kinase : & 417-418 & 0.0031 & -0.0169
& 0.0008 \\
$\bigtriangledown$ & Globulin : & 399-413 & 0.0036 & -0.0277 & 0.0022 \\
\end{tabular}
\renewcommand{\arraystretch}{1.5}
\end{table}
[Keys: m $\to$ slope; c $\to$ constant; sd $\to$ standard deviation]

\newpage
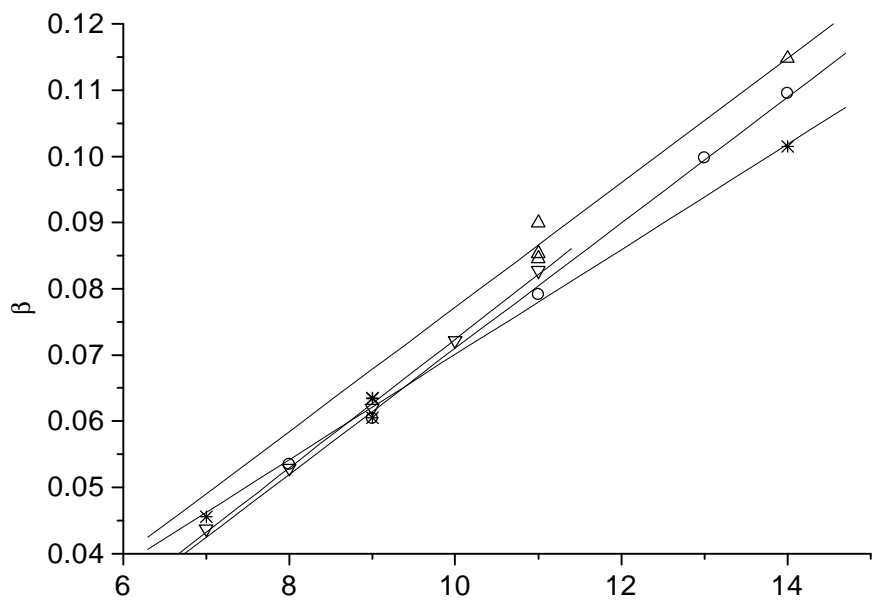
Figure 3. The text-length (l) is kept fixed at 140 to plot $\beta$ as a function of f$_1$ for the CDS of the same 4 proteins as in figure 2. The best fit here is a linear one. 

\begin{table}[h]
\renewcommand{\arraystretch}{1.5}
\begin{tabular}{c c  c c c }
symbol & CDS  & m & c & sd \\
\hline
$\star$ & $\alpha$-globin : & 0.0080 & -0.0093 & 0.0015 \\
$\circ$ & $\beta$-globin  : & 0.0095 & -0.0239 & 0.0013 \\
$\bigtriangleup$ & phosphoglycerate kinase : & 0.0094 & -0.0167
& 0.0029 \\
$\bigtriangledown$ & Globulin :  & 0.0097 & -0.0250 & 0.0007 \\
\end{tabular}
\renewcommand{\arraystretch}{1.5}
\end{table}
[Keys: m $\to$ slope; c $\to$ constant; sd $\to$ standard deviation]

\newpage
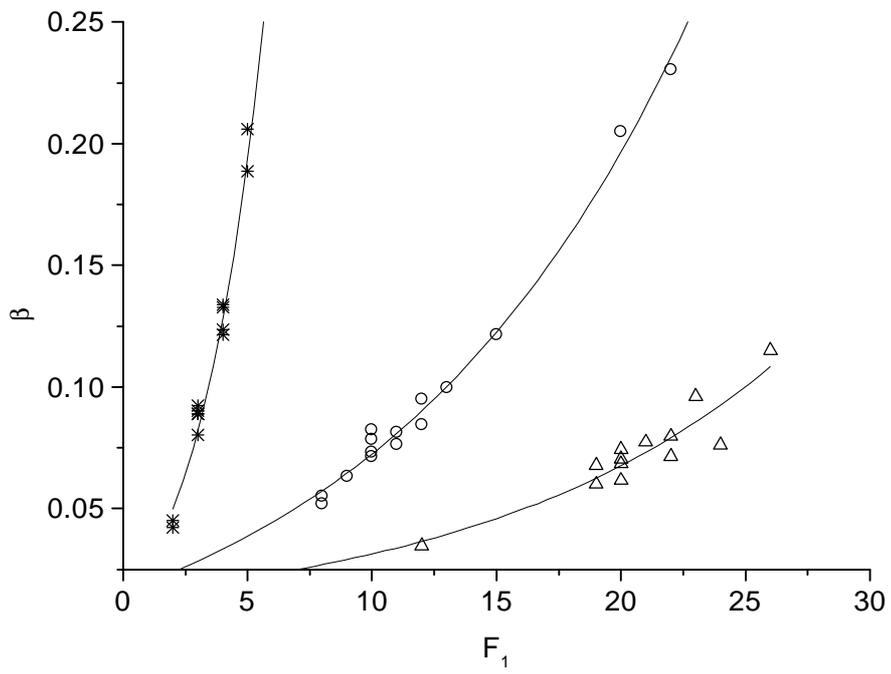
Figure 4. $\beta$ is plotted as a function of f$_1$ at 3 different values of l. Here a number of different CDS from various species are chosen and cut at 3 text-lengths 30, 140 and 300. For text-lengths 30 and 140, 15 CDS were chosen (GenBank  accession numbers are AF007570, L37416, M16024, AF053332, AF001310, M15387, V00410, M15052, L47295, X07083, M59772, J05118, AF056080, AF170848 and M64656), while for text-length 300, 13 CDS were chosen (GenBank  accession numbers are U02504, AF000953, M73993, AF054895, AF076528, AF053332, M15052, U65090, Z54364, U53218, AB013732, M15668 and U69698). Unlike figure 2 and figure 3, the exponential gives the better fit over the linear. The fit function: Y=Y0 + A.e$^{(X/t)}$ .

\begin{table}[h]
\renewcommand{\arraystretch}{1.5}
\begin{tabular}{c c c c c }
symbol & l & Y0 & A & t \\
\hline
$\star$ & 30 & 0.0236 & 0.0357 & 2.7704 \\
$\circ$ & 140 & 0.0324 & 0.0481 & 12.8086 \\
$\bigtriangleup$ & 300 & 0.0018 & 0.0133 & 12.4689 \\
\end{tabular}
\renewcommand{\arraystretch}{1.5}
\end{table}

\newpage
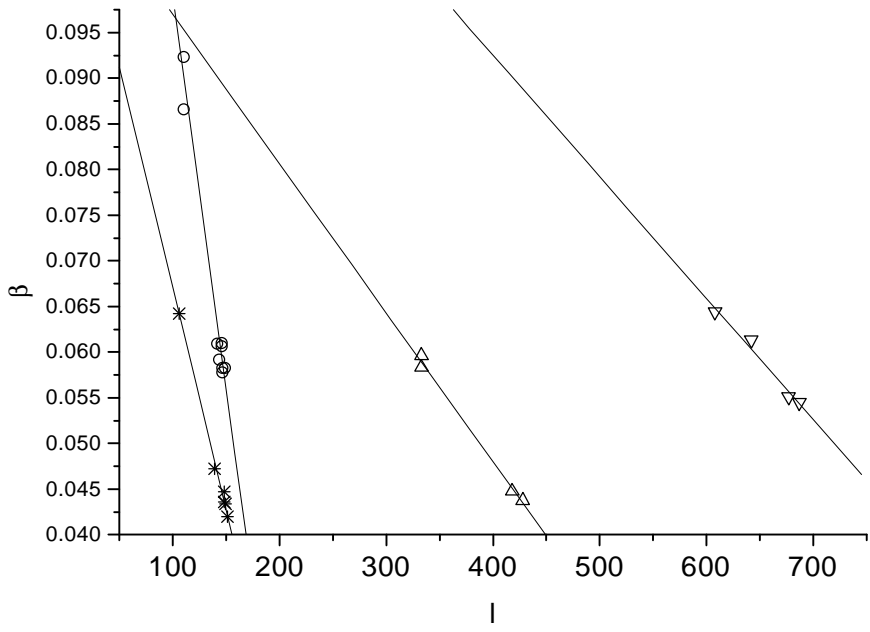
Figure 5. $\beta$ is plotted as a function of l for 4 different values of f$_1$. For each f$_1$, natural CDS of that particular f$_1$, are considered. The relationship between $\beta$ and l for fixed f$_1$ comes out to be linear. 

\begin{table}[h]
\renewcommand{\arraystretch}{1.5}
\begin{tabular}{c c c c c }
symbol & f$_1$ & m & c & sd \\
\hline
$\star$ & 7 & -4.84$\times$10$^{-4}$ & 0.1154 & 6.89$\times$10$^{-4}$ \\
$\circ$ & 9 & -8.54$\times$10$^{-4}$ & 0.1841 & 0.0021 \\
$\bigtriangleup$ & 20 & -1.63$\times$10$^{-4}$ & 0.1133 & 7.14$\times$10$^{-4}$ \\
$\bigtriangledown$ & 38 & -1.33$\times$10$^{-4}$ & 0.1458 & 8.85$\times$10$^{-4}$ \\
\end{tabular}
\renewcommand{\arraystretch}{1.5}
\end{table}
[Keys: m $\to$ slope; c $\to$ constant; sd $\to$ standard deviation]

\newpage
\begin{table}
\caption{\bf{The $\beta$ values for some CDS from different organisms. The l and $f_{1}$ stand for the total number of the triplet codons and the frequency of the most frequent codon respectively.  The $\chi^{2}$ value signifies how good the fit is and the degrees of freedom, denoted by df, is simply one less than the total number of ranks. The $\beta_{Th}$ and $\beta^{'}$ are explained in equation (10).}} 
\bigskip
\renewcommand{\arraystretch}{1.5}
\begin{tabular}{|c|c|c|c|c|c|c|c|c|c|}
\hline
Protein & Organism & Accession no. & l & $f{_1}$ & $\beta$ & $\chi^{2}$ 
& df & $\beta_{Th}$ & $\beta^{'}$  \\
\hline
$\alpha$-globin & Ark Clam & X71386 & 151 & 7 & 0.04221 & 0.137 & 52 & 0.0405 
& 1.0415 \\
& Rainbow Trout & D88114 & 144 & 9 & 0.05893 & 0.202 & 43 & 0.0571
& 1.0321  \\
& $\it{Cyprinus \: carpio}$ & AB004739 & 144 & 10 & 0.06890 & 0.450 & 45 
& 0.0645 & 1.0691 \\
& Black Rockcod & AF049916 & 144 & 11 & 0.07649 & 0.594 & 41 & 0.0719
& 1.0646 \\
& Duck & J00923 & 143 & 10 & 0.06801 & 0.105 & 40 & 0.0645 & 1.0551 \\
& Pigeon & X56349 & 143 & 10 & 0.06872 & 0.155 & 40 & 0.0649 & 1.0584 \\
\hline
\end{tabular}
\renewcommand{\arraystretch}{1.5}
\end{table}

\newpage
\begin{table}
\renewcommand{\arraystretch}{1.5}
\begin{tabular}{|c|c|c|c|c|c|c|c|c|c|}
\hline
Protein & Organism & Accession no. & l & $f{_1}$ & $\beta$ & $\chi^{2}$ 
& df & $\beta_{Th}$ & $\beta^{'}$ \\
\hline
$\alpha$-globin & Chicken & V00410 & 142 & 10 & 0.07251 & 0.893 & 46 & 0.0654 & 1.1089 \\
& House Mouse & V00714 & 142 & 9 & 0.06037 & 0.192 & 45 & 0.0579 & 1.0421 \\
& Rhesus Monkey & J004495 & 143 & 10 & 0.06568 & 0.353 & 37 & 0.0649
& 1.0117 \\
& Rabbit & M11113 & 143 & 10 & 0.06661 & 0.188 & 38 & 0.0649 & 1.0260 \\
& Norway Rat & U62315 & 143 & 10 & 0.06897 & 0.386 & 43 & 0.0649 & 1.0624 \\
& Otolemur & M29648 & 143 & 13 & 0.09286 & 0.727 & 38 & 0.0874 & 1.0620 \\
& Grevy's Zebra & U70191 & 143 & 13 & 0.09678 & 0.272 & 40 & 0.0874 
& 1.1068 \\
& Human & V00488 & 143 & 14 & 0.10045 & 0.007 & 35 & 0.0950 & 1.0569 \\
& Orangutan & M12157 & 143 & 15 & 0.11022 & 0.487 & 37 & 0.1027 & 1.0732 \\
& Horse & M17902 & 143 & 15 & 0.11385 & 0.399 & 40 & 0.1027 & 1.1086 \\
\hline
\end{tabular}
\renewcommand{\arraystretch}{1.5}
\end{table}

\newpage
\begin{table}
\renewcommand{\arraystretch}{1.5}
\begin{tabular}{|c|c|c|c|c|c|c|c|c|c|}
\hline
Protein & Organism & Accession no. & l & $f{_1}$ & $\beta$ & $\chi^{2}$ 
& df & $\beta_{Th}$ & $\beta^{'}$ \\
\hline
$\alpha$-globin & Sheep & X70215 & 143 & 17 & 0.13269 & 1.153 & 38 & 0.1182 
& 1.1231 \\
& Goat & J00043 & 143 & 17 & 0.13675 & 1.432 & 41 & 0.1182 & 1.1574 \\
& Salamander & M13365 & 144 & 9 & 0.06240 & 0.489 & 51 & 0.0571 & 1.0928 \\
& Clawed Frog & X14260 & 142 & 10 & 0.07394 & 0.411 & 48 & 0.0654 & 1.1308 \\
\hline
$\beta$-globin & Atlantic Salmon & X69958 & 149 & 11 & 0.07382 & 0.543 
& 43 & 0.0694 & 1.0643 \\
& Clawed Frog & Y00501 & 147 & 9 & 0.05772 & 0.196 & 45 & 0.0559 & 1.0326 \\
& Chicken & V00409 & 147 & 9 & 0.05773 & 0.324 & 46 & 0.0559 & 1.0327 \\
& House Mouse & V00722 & 147 & 8 & 0.05075 & 0.099 & 46 & 0.0488 & 1.0410 \\
& Rabbit & V00882 & 146 & 9 & 0.06091 & 0.133 & 46 & 0.0563 & 1.0817 \\
& Rat & X06701 & 147 & 10 & 0.06849 & 0.545 & 43 & 0.0631 & 1.0856 \\
\hline
\end{tabular}
\renewcommand{\arraystretch}{1.5}
\end{table}

\newpage
\begin{table}
\renewcommand{\arraystretch}{1.5}
\begin{tabular}{|c|c|c|c|c|c|c|c|c|c|}
\hline
Protein & Organism & Accession no. & l & $f{_1}$ & $\beta$ & $\chi^{2}$ 
& df & $\beta_{Th}$ & $\beta^{'}$ \\
\hline
$\beta$-globin & Oppossum & J03643 & 148 & 12 & 0.08164 & 2.183 & 45 & 0.0771 & 1.0592 \\
& Sheep & X14727 & 146 & 12 & 0.08413 & 0.351 & 39 & 0.0782 & 1.0761 \\
& Goat & M15387 & 146 & 13 & 0.09558 & 0.406 & 42 & 0.0856 & 1.1170 \\
& Lemur & M15734 & 148 & 14 & 0.10743 & 1.375 & 42 & 0.0917 & 1.1715 \\
& Human & AF007546 & 148 & 15 & 0.11245 & 1.530 & 39 & 0.0991 & 1.1349 \\
\hline
Insulin & Salmon & J00936 & 106 & 7 & 0.06425 & 0.490 & 45 & 0.0582 
& 1.1040 \\
& Clawed Frog & M24443 & 107 & 8 & 0.07922 & 0.841 & 46 & 0.0676 & 1.1726 \\
& Syrian Hamster & M26328 & 111 & 9 & 0.08656 & 0.703 & 42 & 0.0747 
& 1.1592 \\
& Guinea Pig & K02233 & 111 & 9 & 0.09220 & 0.815 & 45 & 0.0747 & 1.2348 \\
& Owl Monkey & J02989 & 109 & 13 & 0.14189 & 1.667 & 39 & 0.1162 & 1.2216 \\
& $\it{Octodon \: degus}$ & M57671 & 110 & 12 & 0.14122 & 1.322 & 44 
& 0.1050 & 1.345 \\
\hline
\end{tabular}
\renewcommand{\arraystretch}{1.5}
\end{table}

\newpage
\begin{table}
\renewcommand{\arraystretch}{1.5}
\begin{tabular}{|c|c|c|c|c|c|c|c|c|c|}
\hline
Protein & Organism & Accession no. & l & $f{_1}$ & $\beta$ & $\chi^{2}$ 
& df & $\beta_{Th}$ & $\beta^{'}$ \\
\hline
Insulin & Rat & J00747 & 111 & 12 & 0.14785 & 2.192 & 44 & 0.1040 & 1.4216 \\
& Human & J00265 & 111 & 13 & 0.17379 & 2.795 & 42 & 0.1240 & 1.4012 \\
& Rabbit & U03610 & 111 & 18 & 0.21253 & 2.940 & 32 & 0.1648 & 1.2890 \\
\hline
Globulin & Pig & AF204929 & 413 & 18 & 0.03901 & 0.860 & 58 & 0.0420 
& 0.9286 \\
& Bovine & AF204928 & 412 & 19 & 0.04173 & 1.227 & 57 & 0.0446 & 0.9348 \\
& Djungarian Hamster & U16673 & 400 & 25 & 0.06195 & 5.871 & 59 & 0.0618
& 1.0024 \\
& Norway Rat & NM$\_$012650 & 404 & 26 & 0.06505 & 7.256 & 59 & 0.0638
& 1.0196 \\
& House Mouse & NM$\_$011367 & 404 & 28 & 0.07215 & 9.484 & 58 & 0.0691
& 1.0447 \\
& Human & NM$\_$001040 & 403 & 33 & 0.09463 & 18.202 & 60 & 0.1112 & 0.8511\\
& Rabbit & AF144711 & 399 & 39 & 0.12568 & 19.189 & 60 & 0.0998 & 1.2596 \\ 
\hline 
\end{tabular}
\renewcommand{\arraystretch}{1.5}
\end{table}

\newpage
\begin{table}
\renewcommand{\arraystretch}{1.5}
\begin{tabular}{|c|c|c|c|c|c|c|c|c|c|}
\hline
Protein & Organism & Accession no. & l & $f{_1}$ & $\beta$ & $\chi^{2}$ 
& df & $\beta_{Th}$ & $\beta^{'}$ \\
\hline
Heat shock & $\it{Babesia \: microti}$ & U53448 & 646 & 35 & 0.05127 & 0.867 & 55 & 0.0540 & 0.9491 \\ 
protein 70 & Pacific Oyster & AF144646 & 660 & 36 & 0.05235 & 1.576 & 58 
& 0.0544 & 0.9616 \\
& Human & U56725 & 640 & 40 & 0.06454 & 3.140  & 59 & 0.0628
& 1.0277 \\
& Mouse & L27086 & 642 & 38 & 0.06131 & 2.627  & 60 & 0.0593 & 1.0341 \\
& Chinook Salmon & U35064 & 645 & 42 & 0.06640 & 1.533 & 60 & 0.06559
& 1.0124 \\
& Rat & L16764 & 642 & 48 & 0.07369 & 6.523 & 40 & 0.0759 & 0.9710 \\
\hline
Phospho- & Human & X80497 & 1236 & 51 & 0.03709 & 10.391 & 61 
& 0.0413 & 0.8990  \\
rylase & Rabbit & X60421 & 1236 & 58 & 0.04458 & 7.694 & 61 & 0.0472
& 0.9449 \\
kinase & Mouse & X74616 & 1242 & 47 & 0.03244 & 8.927 & 61 & 0.0377 & 0.8598 \\
\hline
Glycogen & Human & J04501 & 738 & 44 & 0.05968 & 6.984 & 60 & 0.0599
& 0.9952 \\
synthase & Mouse & U53218 & 739 & 37 & 0.04718 & 7.113 & 60 & 0.0499
& 0.9455 \\
\hline 
\end{tabular}
\renewcommand{\arraystretch}{1.5}
\end{table}

\newpage
\begin{table}
\renewcommand{\arraystretch}{1.5}
\begin{tabular}{|c|c|c|c|c|c|c|c|c|c|}
\hline
Protein & Organism & Accession no. & l & $f{_1}$ & $\beta$ & $\chi^{2}$ 
& df & $\beta_{Th}$ & $\beta^{'}$ \\
\hline
Glycogen& Rabbit & AF017114 & 736 & 49 & 0.06603 & 3.001 & 59 & 0.0674 
& 0.9804 \\
synthase & Rat & J05446 & 704 & 28 & 0.03483 & 1.945 & 60 & 0.0391 
& 0.8910 \\
\hline
Troponin C & Chicken & M16024 & 162 & 17 & 0.12374 & 1.577 & 45 & 0.1037
& 1.1938 \\
& Human & M22307 & 161 & 23 & 0.19581 & 3.333 & 40 & 0.1460 & 1.3413 \\
& Mouse  & M57590 & 161 & 21 & 0.17806 & 4.565 & 42 & 0.1319 & 1.3496 \\
& Rabbit & J03462 & 161 & 24 & 0.19294 & 3.964 & 36 & 0.1531 & 1.2606 \\
& Clawed Frog & AB003080 & 162 & 16 & 0.12250 & 1.370 & 47 & 0.0969 
& 1.2645 \\
\hline
Albumin & Bovine & M73993 & 608 & 38 & 0.06437 & 9.754 & 59 & 0.0627
& 1.0265 \\
& Human & NM${\_}$001133 & 600 & 34 & 0.05643 & 9.235 & 58 & 0.0565
& 0.9986 \\
& Clawed Frog & M18350 & 607 & 41 & 0.06845 & 15.699 & 56 & 0.0681
& 1.0056 \\
\hline 
\end{tabular}
\renewcommand{\arraystretch}{1.5}
\end{table}

\newpage
\begin{table}
\renewcommand{\arraystretch}{1.5}
\begin{tabular}{|c|c|c|c|c|c|c|c|c|c|}
\hline
Protein & Organism & Accession no. & l & $f{_1}$ & $\beta$ & $\chi^{2}$ 
& df & $\beta_{Th}$ & $\beta^{'}$ \\
\hline
Lysozyme & $\it{Anopheles \: gambiae}$ & U28809 & 141 & 11 & 0.08073 & 0.561 & 45 & 0.0734 & 1.0993 \\
& Bovine & M95099 & 148 & 7 & 0.04359 & 0.094 & 51 & 0.0414 & 1.0539 \\
& $\it{Cyprinus \: carpio}$ & AB027305 & 146 & 9 & 0.06056 & 0.390 & 47 
& 0.0563 & 1.0757 \\
& Human & M19045 & 149 & 7 & 0.04341 & 0.122 & 52 & 0.0411 & 1.0567 \\
& Pig & U44435 & 149 & 8 & 0.04946 & 0.503 & 51 & 0.0481 & 1.0287 \\
\hline
Lactate  & Alligator & L79952 & 334 & 16 & 0.05460 & 0.441 & 58 & 0.0459
& 1.1890 \\
dehydro- & $\it{Cyprinus \: carpio}$ & AF076528 & 334 & 23 & 0.0708 
& 2.166 & 53 & 0.0680 & 1.0401 \\
genase & Human & U13680 & 333 & 20 & 0.05961 & 3.075 & 57 & 0.0587 & 1.0157 \\
&  Pig & U95378 & 333 & 19 & 0.05461 & 2.347 & 57 & 0.0555 & 0.9838 \\
& Pigeon & L79957 & 334 & 19 & 0.05536 & 2.110 & 56 & 0.0553 & 1.0003 \\
& Clawed Frog & AF070953 & 333 & 20 & 0.05831 & 2.010 & 53 & 0.0586
& 0.9935 \\
\hline
\end{tabular}
\renewcommand{\arraystretch}{1.5}
\end{table}

\newpage
\begin{table}
\renewcommand{\arraystretch}{1.5}
\begin{tabular}{|c|c|c|c|c|c|c|c|c|c|}
\hline
Protein & Organism & Accession no. & l & $f{_1}$ & $\beta$ & $\chi^{2}$ 
& df & $\beta_{Th}$ & $\beta^{'}$ \\
\hline
Phospho- & $\it{Candida \:albicans}$ & U25180 & 418 & 34 & 0.08126 & 2.388 
& 38 & 0.0821 & 0.9901 \\
glycerate & $\it{Leishmania \: major}$ & L25120 & 418 & 34 & 0.08677 
& 1.132 & 56 & 0.0821 & 1.0573 \\
kinase & Mouse & M15668 & 418 & 23 & 0.05298 & 1.155 & 58 & 0.0540 & 0.9807 \\
& Rat & M31788 & 418 & 23 & 0.05374 & 1.825 & 60 & 0.0540 & 0.9948 \\
& $\it{Schistosoma \: mansoni}$ & L36833 & 417 & 29  & 0.07284
& 5.498 & 60 & 0.0694 & 1.0494 \\
\hline
Carboxy- & $\it{Aedes \: aegypti}$ & AF165923 & 428 & 20 & 0.04373 
& 1.785 & 61 & 0.0454 & 0.9636 \\
peptidase & Bovine & M61851 & 420 & 22 & 0.05170 & 0.417 & 59 & 0.0512 & 1.0088 \\
A & Human & M27717 & 418 & 20 & 0.04477 & 1.128 & 59 & 0.0465 & 0.9630 \\
& Mouse & J05118 & 418 & 23 & 0.05124 & 6.547 & 58 & 0.0540 & 0.9485 \\
\hline
\end{tabular}
\renewcommand{\arraystretch}{1.5}
\end{table}

\end{document}